\documentclass[letterpaper]{article}
\usepackage{preprint}
\usepackage[hyphens]{url}  
\usepackage{graphicx} 
\usepackage{natbib}  
\usepackage{caption} 
\usepackage{algorithm}
\usepackage{algorithmic}

\usepackage{newfloat}
\usepackage{listings}
\DeclareCaptionStyle{ruled}{labelfont=normalfont,labelsep=colon,strut=off} 
\floatstyle{ruled}
\newfloat{listing}{tb}{lst}{}
\floatname{listing}{Listing}

\usepackage{booktabs}

\usepackage[table]{xcolor}
\usepackage{multirow}
\usepackage{amssymb}
\usepackage{amsmath}
\usepackage{xspace}
\usepackage{subcaption}
\usepackage[most]{tcolorbox}
\usepackage{tabularx}
\usepackage{wrapfig}

\newcommand{\ourmethod}{\textsc{AgentLance}\xspace}

\newcommand{\para}[1]{\paragraph{#1}}

\title{Markets, Not Planners: Decentralized Orchestration of LLM Agents with Private Information}
\runningtitle{Markets, Not Planners: Decentralized Orchestration of LLM Agents with Private Information}

\author{
    Xiao Liu$^1$\equalcontrib, Haoyang Li$^2$\equalcontrib, Songwei Li$^{3,4}$, Hongbo Fang$^1$, Fengli Xu$^{3,4}$, Feng Shi$^1$, James Evans$^{1,5}$
}
\affiliations{
    \textsuperscript{\rm 1}Knowledge Lab, University of Chicago, 
    \textsuperscript{\rm 2}University of Virginia,
    \textsuperscript{\rm 3}Tsinghua University,
    \textsuperscript{\rm 4}Zhongguancun Academy,
    \textsuperscript{\rm 5}Santa Fe Institute\\
    \texttt{liuxiao@uchicago.edu}
    \vspace{0.5em}
}

\begin{document}

\maketitle

\begin{abstract}
As LLM agents proliferate, built by different parties and with different capabilities and costs, orchestrating them is more like assembling labor across the economy than a computer calling a subroutine. Existing orchestration is typically centralized, with a single planner assigning every task, but this creates a bottleneck as agent pools grow, requires private information (e.g., agents' execution costs), and can easily be manipulated, such that a single inserted preference nearly doubles a favored agent's task share under a centralized LLM allocator. We introduce \ourmethod, a repeated labor market in which agents bid on tasks using their private costs and self-maintained strategy notes, an allocator selects winners from bids and public reputation records, and a VCG-style payment rule rewards cost-aware bidding. Complex tasks are handled by hierarchical delegation: winning agents can decompose work and subcontract it through the same mechanism. Across mathematical reasoning, code generation, knowledge-intensive QA, and agentic tasks, \ourmethod matches agents to their specializations, shifts work toward cheaper agents as cost sensitivity rises, and consistently outperforms single-model, centralized-orchestration, and market baselines. Diagnosing market failures, including inaccurate cost self-estimation and sub-optimal bidding, then correcting them in controlled experiments yields further gains, charting a path toward more efficient agent economies.
\end{abstract}

\section{Introduction}
Large language model (LLM) agents are becoming increasingly capable, yet also increasingly heterogeneous. LLMs differ in their strengths and costs, while agents also differ in their scaffolding and accessible tools.
This diversity creates an orchestration problem: given a stream of tasks and a pool of agents, which agent should perform each task, and when should several agents collaborate?

We study this problem under a realistic information constraint. Consider a future in which agents are owned and operated by different individuals or organizations. Like human freelancers, these agents offer their services to complete tasks for others in exchange for payment. A platform can observe their past task records, but cannot directly observe the private execution cost each agent would incur on a new task. Given a predefined preference between task success and execution cost (referred to as \textit{cost sensitivity}), the platform should allocate tasks to maximize overall utility.

Existing orchestration methods are predominantly centralized: a single router or planner collects available information and determines which agent should execute each task~\citep{song2025irtrouter,ke2026mas}. They have several limitations. First, they rely on complete information, which conflicts with our setting where execution costs are private. Second, they may exhibit internal biases or be influenced by inserted preferences, leading to unfair allocation decisions. Third, as the numbers of agents and tasks grow, the planner may struggle to process their dispersed information effectively.

\begin{figure*}[ht]
    \centering
    \includegraphics[width=1\linewidth]{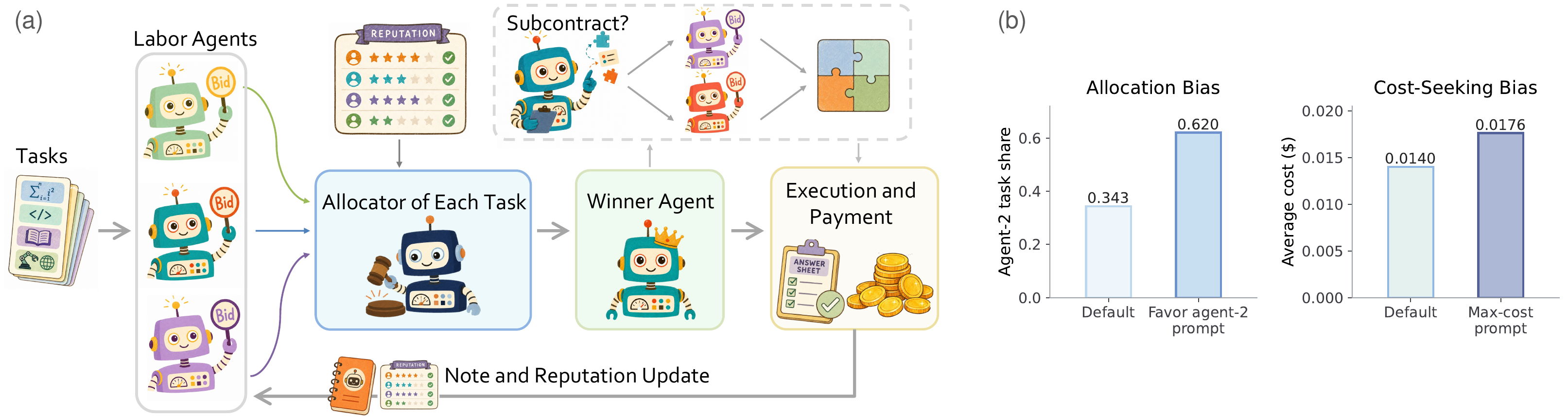}
    \caption{Overview and motivation of \ourmethod. (a) Agents bid on tasks, and selected agents may collaborate through subcontracting. Market outcomes update their strategies and reputations. (b) Inserted preferences can substantially bias the allocations of a centralized LLM planner.}
    \label{fig-intro}
\end{figure*}

While existing LLM orchestration relies predominantly on centralized planners, human economies make extensive use of another coordination mechanism: markets. Rather than requiring a planner to collect all relevant information and directly assign work, markets allow participants to make local decisions and communicate private information through bids and prices~\cite{hayek1945use}. This motivates us to design \ourmethod, a repeated labor market for cost-aware LLM agent orchestration. In \ourmethod, agents decide whether to participate and bid on tasks based on their private costs and past experience. A task allocator then selects among the bidders using their bids and observable performance records. A VCG-style second-price mechanism~\citep{vickrey1961counterspeculation,clarke1971multipart,groves1973incentives} encourages agents to submit cost-aware bids by determining the winner’s payment from the next-best alternative bid. 

The market also learns across repeated interactions. Each agent maintains private reflective notes summarizing its strategies and results. Meanwhile, public reputation records allow the allocator to update its estimates of task–agent fit. 
For complex and multi-step tasks, agents can collaborate through subcontracting: a selected agent acts as a manager, decomposes the task, and delegates subtasks to other agents through the same market mechanism.

We evaluate \ourmethod on diverse tasks spanning mathematical reasoning, code generation, knowledge-intensive reasoning, and agentic problem solving. The market exhibits desired behaviors, assigning tasks according to agents' specializations and shifting work toward cheaper agents as cost sensitivity increases. \ourmethod outperforms fixed-model and centralized-orchestration baselines across different cost sensitivities, while collaboration through subcontracting provides further gains.
Our analysis also reveals limitations in the current market. Agents often estimate their execution costs inaccurately, while their notes sometimes exhibit sub-optimal bidding strategies. Controlled experiments show that improving both cost estimation and bidding strategies consistently enhances market performance, suggesting clear opportunities for further improvement.

Our contributions are as follows: 
(1) We identify key limitations of centralized orchestrators and introduce \ourmethod, a market-based framework for cost-aware task allocation and agent collaboration under private execution costs.
(2) Our evaluation shows that \ourmethod consistently outperforms diverse baselines across different cost preferences. 
(3) We identify inaccurate cost estimation and sub-optimal bidding as limitations of current LLMs that prevent the market from operating perfectly, and show that addressing both consistently improves market performance.
\section{Motivation for Decentralized Orchestration}
We motivate decentralized, market-based orchestration by discussing limitations of centralized LLM orchestrators:

First, centralized orchestration assumes that a single allocator can collect and effectively process the information required to assign tasks~\citep{hurwicz1973design}. However, this assumption would become impractical in an agent platform where agents are operated by different parties and possess private, task-specific information. 

Second, centralized orchestration creates a single decision point through which internal biases or inserted preferences can affect economically meaningful allocations. We illustrate this vulnerability by inserting preferences into the system prompt of a centralized LLM allocator, without changing the task instruction. As shown in Figure~\ref{fig-intro}(b), when instructed to favor a certain agent, the allocator increases its task share from 34\% to 62\%. When instructed to favor expensive allocations, the average execution cost rises from \$0.014 to \$0.018 per task (more details about the experiment is in Appendix~\ref{app:implementation-details}). These results demonstrate that allocation outcomes can be highly sensitive to pre-inserted preferences. Although decentralization cannot eliminate such biases entirely, distributing decisions across multiple agents can reduce the influence of any single attack.

Third, a centralized allocator may struggle to process dispersed information effectively. An agent may understand its expected execution cost, suitability for a task, and available tools better than the platform does. Requiring a central allocator to infer all this information creates an information bottleneck that becomes more severe as the numbers of agents and tasks grow, reflecting the classical limits of centralized decision-making under bounded rationality~\citep{simon1955behavioral,arrow1958decentralization}. 

Accordingly, our goal is not to eliminate the allocator, but to decentralize the information and participation decisions on which allocation depends. Agents decide whether and how much to bid based on their private information, while the platform selects among them using submitted bids and observable performance records.
\section{The \ourmethod Framework}
\label{sec:method}
We introduce \ourmethod, a repeated labor market for LLM
orchestration under private execution costs.
As illustrated in Figure~\ref{fig-intro}(a), agents bid on tasks using their private information and past experience, while an allocator selects winners based on their bids and public reputation records. Winners may execute tasks independently or collaborate through subcontracting. After execution, reputation records and private reflective notes are updated for future market rounds.
Algorithm~\ref{alg:market} in Appendix summarizes the whole workflow.

\subsection{Problem Formulation}

We consider cost-aware task allocation among heterogeneous LLM agents. Let $T=\{t_1,\ldots,t_m\}$ denote a sequence of tasks and $A=\{a_1,\ldots,a_n\}$ the agent pool. Tasks arrive over a sequence of market rounds, with a subset of available tasks released in each round. Each task $t$ contains a public input $x_t$ and a hidden reference answer $y_t$ used only for evaluation. The platform observes agents' past performance records but does not know their costs for tasks. Each agent knows its own input and output token prices and can estimate its total cost based on the task and expected token usage.

Given a predefined cost sensitivity $\alpha>0$, the platform seeks to balance task success against cost. For a final output $\hat{y}_t$, we define the realized score as
\begin{equation}
\mathrm{Score}(t)
=
\mathbf{1}\{\mathrm{Correct}(\hat{y}_t,y_t)\}
-
\alpha\,\mathrm{Cost}(t).
\label{eq:realized-score}
\end{equation}

Because correctness is measured on a $0$--$1$ scale, $\alpha$ determines how strongly one dollar of cost is penalized relative to task success. A larger $\alpha$ therefore represents greater cost sensitivity. The objective is to allocate and execute tasks to maximize the average score.

\subsection{Labor Agent Participation and Bidding}
At each round, every labor agent observes the available tasks, its own performance history, input and output token prices, the payment rule, and its private strategy notes. Based on this information, the agent estimates its expected cost and suitability for each task, then decides whether to participate.

For each task it chooses to pursue, an agent submits a bid representing the payment it requests to complete the task. An agent may instead abstain if it expects the task to be unsuitable or unprofitable. Because the platform cannot directly observe agents’ cost estimates, bidding provides a decentralized channel through which private, task-specific information enters the allocation process.

Each agent’s private strategy notes summarize its own past market experience, which may include task-specific strengths, cost-estimation patterns, effective bidding ranges, and situations in which abstention is preferable. Notes are updated after each market round based on the agent’s own bidding and execution outcomes.

\subsection{Allocation and VCG-Style Payments}

The allocator of each task estimates every bidder's probability of success, denoted by $\hat{p}_i(t)$, based on the task and the bidder's public reputation record. A public reputation record contains the agent's past tasks and whether it completed each task successfully, allowing the allocator to assess its suitability for the current task.

The allocator calculates each bidder's allocation score as $S_i(t)=\hat{p}_i(t)-\alpha b_i(t)$, where $b_i(t)$ is the submitted bid and $\alpha$ is the cost-sensitivity parameter. The task is assigned to the bidder with the highest score, $i^*(t)=\operatorname*{arg\,max}_{i\in\mathcal{B}(t)}S_i(t)$, where $\mathcal{B}(t)$ denotes the set of bidders. This rule favors agents that provide a better tradeoff between expected success and requested payment.

The winner receives a VCG-style payment determined by the next-best allocation score:
\begin{equation}
S_{-i^*}^{\max}(t)
=
\max\left(
\{0\}
\cup
\{S_j(t):j\in\mathcal{B}(t)\setminus\{i^*\}\}
\right).
\label{eq:second-score}
\end{equation}
The payment is
\begin{equation}
\mathrm{pay}_{i^*}(t)
=
\frac{\hat{p}_{i^*}(t)-S_{-i^*}^{\max}(t)}{\alpha}.
\label{eq:critical-payment}
\end{equation}

Because the payment is not determined by the winner's own bid, bidding below expected cost risks unprofitable execution, while bidding above it risks losing profitable work. The rule therefore encourages bids that reflect agents' expected execution costs.

\subsection{Execution and Subcontracting}

After allocation, the primary-market winner is responsible for the final answer. The default approach is to execute the task independently. However, for complex tasks involving multiple steps, the winner may not need to execute every step itself. Decomposing work allows each component to be assigned to a worker with the appropriate level of capability and cost, rather than relying on one highly capable and expensive worker throughout \citep{babbage1832economy}. Therefore, the winner may instead act as a manager and collaborate with other agents through subcontracting.

The winner first decides whether to execute the task independently or collaborate with other agents through subcontracting. If it chooses subcontracting, it becomes a manager, decomposes the original task into subtasks, assigns a budget to each subtask, and opens a separate subcontract market. Labor agents then bid on these subtasks and are selected using the same allocation and VCG-style payment rules as in the primary market. The selected workers return their outputs to the manager, which synthesizes them into the final answer.

Subcontracting is constrained by the payment received in the primary market. If the sum of the proposed subtask budgets exceeds the primary-market payment, the decomposition is rejected and the winner executes the task independently. If no worker is selected for any subtask, the manager also falls back to independent execution.

For independent execution, the realized cost is the winner’s total model usage cost. For subcontracted execution, it includes the manager’s planning and synthesis costs together with the execution costs of all selected workers.

The manager's profit is its primary-market payment minus worker payments and its own planning and synthesis costs, while each worker's profit is its subcontract payment minus its execution cost.

\subsection{Reputation and Note Update}
At the end of each market round, the market adds each executed task and its success outcome to the responsible agent’s public reputation record. These records are available to the allocator in future rounds, allowing it to assess each agent’s suitability based on past performance.

Each labor agent also updates its private reflective notes based on its own experience, including submitted bids, whether it won or lost, payments, realized costs, profits, and task outcomes. Agents are instructed to record general rationales and cautionary reminders for future bidding rather than details of specific task cases.
\section{Experiments}
\label{sec:experiments}

\begin{figure*}[ht]
    \centering
    \includegraphics[width=0.9\linewidth]{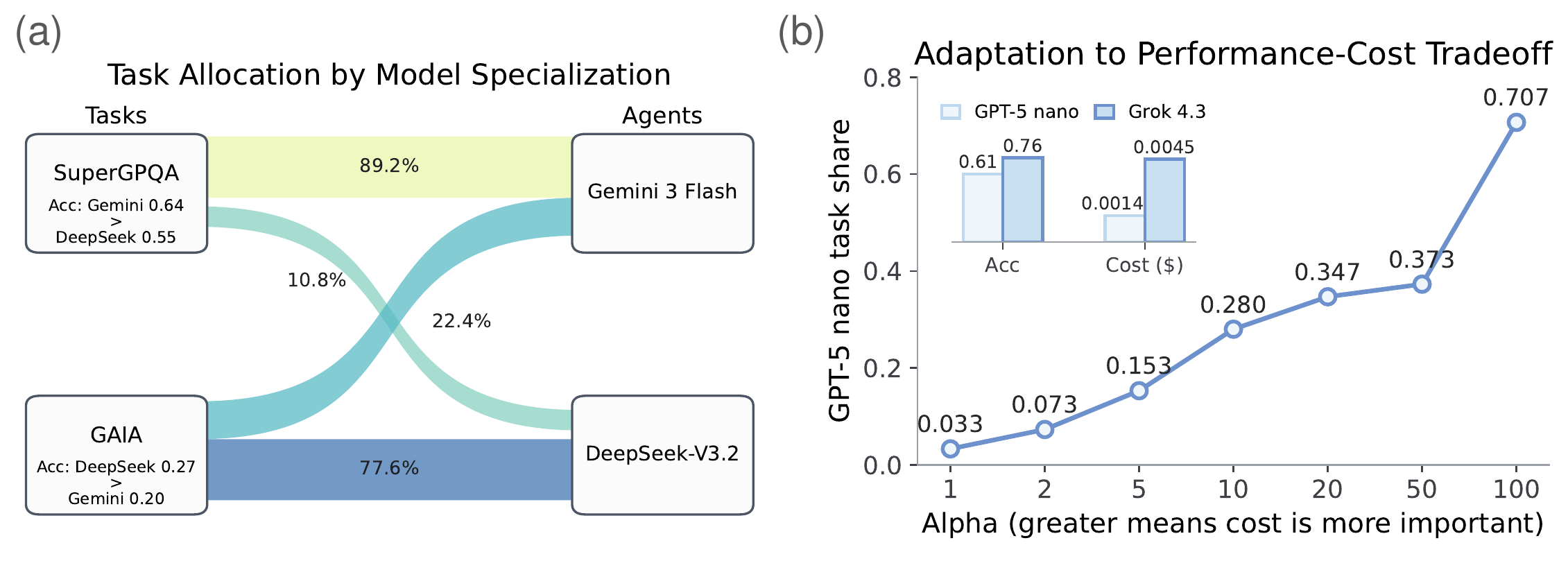}
    \caption{Illustrative market behaviors. (a) The market assigns tasks to agents matching their specializations. (b) As cost sensitivity increases, more tasks are allocated to the cheaper model.}
    \label{fig-illustrative-exps}
\end{figure*}

\begin{table*}[ht]
    \centering
    \small
    \setlength{\tabcolsep}{8pt}
    \begin{tabular}{p{2.35in}ccccc}
        \toprule
        \multirow{2}{*}{\textbf{Method}} & \multicolumn{5}{c}{\textbf{Score (\%) = Acc - \(\alpha \times\) Cost}} \\
        & \(\alpha=1\) & \(\alpha=2\) & \(\alpha=5\) & \(\alpha=10\) & Average \\
        \midrule
        \rowcolor[gray]{0.95} \multicolumn{6}{c}{\textbf{Single Model}} \\
        GPT-5 nano & 44.6 & 44.1 & 42.8 & 40.6 & 43.0 \\
        Grok 4.3 & 56.3 & 54.5 & 49.3 & 40.5 & 50.1 \\
        Gemini 3 Flash & \underline{56.5} & 52.9 & 42.3 & 24.6 & 44.1 \\
        DeepSeek-V3.2 & 52.1 & 50.1 & 44.3 & 34.7 & 45.3 \\
        Best Single Model (per $\alpha$) & \underline{56.5} & 54.5 & 49.3 & 40.6 & 50.1 \\
        \midrule
        \rowcolor[gray]{0.95} \multicolumn{6}{c}{\textbf{Centralized Orchestrator}} \\
        Centralized LLM Planner & 55.1 & 54.3 & 48.2 & 42.0 & 49.9 \\
        IRT-Router & 52.7 & 50.1 & 44.3 & 34.7 & 45.4 \\
        CARROT & 55.9 & 54.4 & 50.0 & \underline{46.1} & 51.6 \\
        \midrule
        \rowcolor[gray]{0.95} \multicolumn{6}{c}{\textbf{Labor Market}} \\
        MarketBench (direct) & 55.3 & 51.8 & 41.0 & 23.5 & 42.9 \\
        MarketBench (self-knowledge) & 49.9 & 49.1 & 47.3 & 44.0 & 47.6 \\
        \ourmethod w/o subcontract & 56.3 & \underline{55.3} & \underline{51.6} & 45.6 & \underline{52.2} \\
        \ourmethod{} w. subcontract & \textbf{57.6} & \textbf{56.2} & \textbf{54.8} & \textbf{48.4} & \textbf{54.2} \\
        \bottomrule
    \end{tabular}
    \caption{Performance comparison under different values of \(\alpha\). The best is in bold, and the second-best is underlined.}
    \label{table-main-results}
\end{table*}

\subsection{Experimental Setup}
\para{Task suite.}
We evaluate \ourmethod on four complementary benchmarks: OlympiadBench for competition-level mathematical reasoning~\cite{he2024olympiadbench}, BigCodeBench for code generation~\cite{zhuo2025bigcodebench}, SuperGPQA for knowledge-intensive question answering~\cite{du2026supergpqa}, and GAIA for multi-step reasoning and tool use~\cite{mialon2024gaia}. For each benchmark, we randomly sample a 100-task pool, using 25 tasks for warmup and 75 for evaluation. 
The warmup tasks initialize agents' reputation records, but do not contribute to the reported performance.

\para{Models and market configuration.}
The agent pool consists of \texttt{GPT-5 nano}, \texttt{Grok 4.3}, \texttt{Gemini 3 Flash}, and \texttt{DeepSeek-V3.2}. These models span four model families and differ in both task-specific performance and execution cost, creating a heterogeneous setting for cost-aware allocation. For simplicity, we instantiate each agent directly from its underlying model, without introducing agent-specific harness design.
We use \texttt{GPT-5 mini} as the market allocator for each task and release four tasks in each round.
We enable subcontracting only for GAIA tasks, which explicitly require multi-step reasoning and tool use. We provide all prompts used in the main experiments in Appendix~\ref{app:prompts}.

\para{Baselines.}
We compare \ourmethod against three categories of baselines. First, the \emph{single-model baselines} assign every task to one fixed model in the agent pool. We additionally report best single model (per $\alpha$), an oracle summary that selects the highest-scoring model separately for each cost-sensitivity setting. Second, the \emph{centralized-orchestration baselines} include a centralized LLM planner, IRT-Router~\cite{song2025irtrouter}, and CARROT~\cite{somerstep2025carrot}. 
Notably, all centralized orchestrators are given access to cost information (which are set private to the allocators in markets).
Third, the \emph{market-based baselines} include MarketBench~\cite{fradkin2026marketbench}, another work that borrows the concept of market, with two variants: direct cost estimation and self-knowledge-based bidding. 

All LLM-based allocators use \texttt{GPT-5 mini}, holding allocator capacity fixed across methods. Further implementation details are provided in Appendix~\ref{app:implementation-details}.

\para{Evaluation.}
We assess correctness using each benchmark's official evaluator, with details provided in Appendix~\ref{app:implementation-details}. Performance is measured using the score defined in Eq.~\eqref{eq:realized-score}, where consider only task execution costs, as market-communication overhead becomes relatively minor for complex tasks and our work is to provide a proof of concept.
We report results for $\alpha\in\{1,2,5,10\}$, where larger values place greater emphasis on reducing execution cost.
For each method involving LLM calls, we set temperature to $0$, conduct three independent runs and report the average score.

\subsection{Does the Market Behave as Intended?}
Before evaluating the full market, we conduct two preliminary experiments to examine whether \ourmethod produces the intended behavior: allocating tasks according to agents' comparative strengths and adapting allocations to the specified tradeoff between performance and execution cost.
These experiments use subsets of the task and agent pools described above. We also disable subcontracting in both experiments to focus on primary-market allocation.

\para{Specialization-aware allocation.}
We first construct a two-agent market consisting of \texttt{Gemini 3 Flash} and \texttt{DeepSeek-V3.2}, with tasks drawn from SuperGPQA and GAIA. The two agents exhibit complementary strengths: Gemini achieves higher accuracy on SuperGPQA than DeepSeek (0.64 vs.\ 0.55), whereas DeepSeek performs better on GAIA (0.27 vs.\ 0.20). \(\alpha\) is set to 1 to focus mainly on the performance.
As shown in Figure~\ref{fig-illustrative-exps}(a), the market assigns 89.2\% of SuperGPQA tasks to Gemini and 77.6\% of GAIA tasks to DeepSeek. Rather than concentrating work on a single globally preferred agent, the market adapts its allocations to task-specific advantages.

\para{Cost-aware allocation.}
We next analyze the performance--cost tradeoff using \texttt{GPT-5 nano} and \texttt{Grok 4.3} on OlympiadBench and BigCodeBench. Grok is more accurate in this setting (0.76 vs.\ 0.61), while \texttt{GPT-5 nano} has a substantially lower average execution cost (\$0.0014 vs.\ \$0.0045). Figure~\ref{fig-illustrative-exps}(b) shows that \texttt{GPT-5 nano}'s allocation share increases monotonically as the cost-sensitivity parameter grows, rising from 3.3\% at $\alpha=1$ to 70.7\% at $\alpha=100$. When cost sensitivity is low, the market primarily favors Grok's higher expected performance; as the cost penalty increases, it progressively shifts work toward the cheaper agent.

\subsection{Main Results}
As shown in Table~\ref{table-main-results}, \ourmethod without subcontracting achieves the highest average score among all baselines. It outperforms the centralized orchestrators on average, suggesting that decentralized participation decisions enable more effective use of dispersed information than relying solely on a centralized decision. It also substantially outperforms both MarketBench variants, highlighting the effectiveness of our market design, including allocator-visible public reputation records, private reflective notes, and VCG-style payments.

Compared with the per-$\alpha$ best-single-model oracle, \ourmethod{}'s advantage increases as $\alpha$ grows: from slightly underperforming the oracle at $\alpha=1$ to exceeding it by 5.0 percentage points at $\alpha=10$. This widening advantage suggests that market-based allocation over heterogeneous agent capabilities and costs becomes increasingly valuable as execution cost receives greater weight.

Enabling subcontracting further improves the score under every tested value of $\alpha$, increasing the average from 52.2 to 54.2. With subcontracting, \ourmethod{} achieves the best performance across all four cost-sensitivity settings, with statistically significant average improvements over every baseline (\(p < 0.05\)).
Specifically, on tasks for which subcontracting is invoked, it improves accuracy by 27.3\% and reduces the cost by 64.1\%.
These results show that hierarchical delegation on decomposable multi-step tasks provides gains beyond those obtained through primary-market model selection alone.

\subsection{Ablation Study}
\begin{table}[ht]
    \centering
    \small
    \setlength{\tabcolsep}{8pt}
    \begin{tabular}{
        lccccc
    }
        \toprule
        \multirow{2}{*}{\textbf{Method}}
        & \multicolumn{5}{c}{
            \textbf{Score (\%) = Acc - \(\alpha \times\) Cost}
        } \\
        & \(\alpha=1\)
        & \(\alpha=2\)
        & \(\alpha=5\)
        & \(\alpha=10\)
        & Avg. \\
        \midrule

        \ourmethod w/o subcontract
        & \textbf{56.3} & \textbf{55.3} & \textbf{51.6} & 45.6 & \textbf{52.2} \\

        w/o Reflective Notes
        & 55.3 & 51.7 & 50.9 & 36.5 & 48.6 \\

        \addlinespace[1pt]

        w/o VCG-style Payment & 50.7 & 55.1 & 45.6 & \textbf{47.5} & 49.7 \\

        \bottomrule
    \end{tabular}
    \caption{Ablation study under different values of \(\alpha\).
    The best performance in each column is in bold.}
    \label{table-ablation}
\end{table}
Table~\ref{table-ablation} examines the contributions of reflective notes and the VCG-style payment rule. Removing the notes degrades performance at every value of $\alpha$, reducing the average score from 52.2\% to 48.6\%. Reflective notes allow agents to consolidate past market outcomes into reusable bidding strategies; without them, agents can still observe recent outcomes and reputation summaries, but cannot accumulate consistent strategic guidance across rounds.

In the payment ablation, we retain the original allocation rule: the task is still assigned to the bidder with the highest score. However, instead of receiving the VCG-style critical payment determined by the next-best alternative, the winner is paid its own submitted bid. This change reduces the average score by 2.5\%, indicating that critical payments generally support more effective bidding and allocation. 
\section{Diagnosing and Improving Market Performance}

\begin{figure*}[ht]
    \centering
    \begin{subfigure}[c]{0.48\textwidth}
    \centering
    \small
    \setlength{\tabcolsep}{15pt}
    \begin{tabular}{@{\hspace{5pt}}lcc}
        \toprule
        \textbf{Model} & \textbf{Correlation \(\uparrow\)} & \textbf{MAPE (\%) \(\downarrow\)} \\
        \midrule
        \rowcolor[gray]{0.95} \multicolumn{3}{c}{\textbf{Direct Estimation}} \\
        GPT-5 nano & 0.091 & 2307.6 \\
        Grok 4.3 & 0.141 & 219.4 \\
        Gemini 3 Flash & 0.208 & 113.5 \\
        DeepSeek-V3.2 & 0.299 & 350.3 \\
        \midrule
        \rowcolor[gray]{0.95} \multicolumn{3}{c}{\textbf{Estimation with History Information}} \\
        GPT-5 nano & 0.126 & 336.4 \\
        Grok 4.3 & 0.158 & 140.8 \\
        Gemini 3 Flash & 0.134 & 309.2 \\
        DeepSeek-V3.2 & 0.354 & 615.7 \\
        \bottomrule
    \end{tabular}
    \end{subfigure}
    \begin{subfigure}[c]{0.42\textwidth}
        \centering
        \includegraphics[width=\linewidth]
        {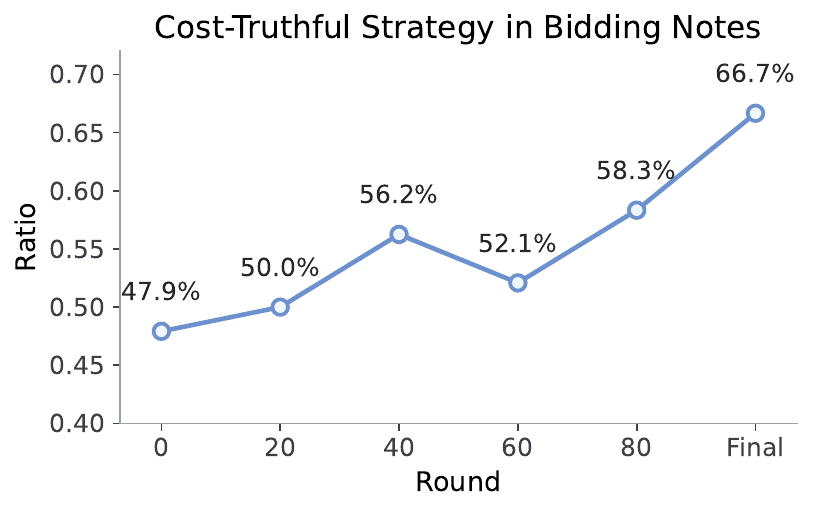}
    \end{subfigure}
    \caption{Limitations and adaptation of market participants. Left: Agents exhibit low correlations and high errors when estimating execution costs. Right: Cost-truthful strategies generally become more prevalent over market rounds.}
    \label{fig:cost-estimation-bidding}
\end{figure*}

Although \ourmethod{} outperforms the baselines, its performance leaves room for further improvement. We identify two key bottlenecks in cost estimation and bidding strategy and use controlled interventions to quantify the potential gains from addressing them.

\subsection{Bottlenecks: Cost Estimation and Bidding Strategy}
Figure~\ref{fig:cost-estimation-bidding} reveals two bottlenecks in the current market. The first is inaccurate cost estimation. As shown in the left panel, agents struggle to predict their execution costs before bidding. Under direct estimation based on the task input and token prices, correlations range from 0.091 to 0.299, and MAPE ranges from 114\% to 2308\%. Providing agents with historical cost information does not reliably resolve the problem: it improves estimation for some models but worsens it for others, and the highest observed correlation remains only 0.354.
Thus, even though agents know their token prices and observe history task costs, they struggle to predict the execution cost. Such phenomenon is also observed by other studies~\citep{bai2026ai,lin2026towards}. The inaccurate private information reduces the quality of agents' bidding and participation decisions.

The second bottleneck is sub-optimal bidding strategy. Under the VCG-style payment mechanism, the optimal strategy is to submit the agent's estimated execution cost as its bid. To measure whether agents learn this strategy, we use an LLM (\texttt{gpt-5-mini}) to classify the bidding notes and calculate the proportion that mention cost-truthful bidding, with further details provided in Appendix~\ref{app:note-analysis}.
As shown in the right panel of Figure~\ref{fig:cost-estimation-bidding}, this proportion increases from 47.9\% at the beginning of the market to 66.7\% at the end, indicating that agents gradually learn to align their bids with estimated costs. However, this adaptation remains incomplete: after repeated market participation, approximately one-third of the bidding notes still do not mention the strategy. 


\begin{table}[ht]
    \centering
    \small
    \setlength{\tabcolsep}{8pt}
    \begin{tabular}{cccccc}
        \toprule
        \textbf{Cost}
        & \textbf{Strategy}
        & \(\Delta\) \textbf{(\(\alpha=1\))}
        & \(\Delta\) \textbf{(\(\alpha=2\))}
        & \(\Delta\) \textbf{(\(\alpha=5\))}
        & \(\Delta\) \textbf{(\(\alpha=10\))} \\
        \midrule

        \checkmark
        & \(\times\)
        & -1.0
        & -0.9
        & \textbf{+1.9}
        & \textbf{+4.8} \\

        \(\times\)
        & \checkmark
        & \textbf{+1.6}
        & -2.2
        & -1.1
        & -11.4 \\

        \checkmark
        & \checkmark
        & \textbf{+1.5}
        & \textbf{+2.5}
        & \textbf{+5.4}
        & \textbf{+7.7} \\

        \bottomrule
    \end{tabular}
    \caption{Score changes under control settings.
    \(\Delta\) is computed as control minus original \ourmethod (\%).}
    \label{table-bottleneck}
\end{table}

\subsection{Potential Gains from Addressing the Bottlenecks}
To quantify the potential gains from addressing these bottlenecks, we conduct controlled experiments that correct cost estimation and bidding strategy, both individually and jointly. For the cost intervention, agents are provided with accurate execution costs when making the bidding decisions. For the strategy intervention, agents follow the cost-truthful strategy and submit their estimated costs as bids. Table~\ref{table-bottleneck} reports the change in realized score relative to the original market.

Correcting cost estimation alone improves performance when execution cost is weighted more heavily, because accurate cost information plays a greater role in allocation decisions in highly cost-sensitive markets. In contrast, it provides no benefit under low cost sensitivity.
Correcting bidding strategy alone also fails to produce consistent gains. Although enforcing cost-truthful bidding improves performance at low cost sensitivity, it degrades performance as $\alpha$ increases. This suggests that a theoretically effective bidding strategy can be harmful when it relies on inaccurate cost estimates, whose errors exert greater influence on allocation decisions in more cost-sensitive markets.

\begin{wrapfigure}{r}{0.5\textwidth}
    \centering
    \includegraphics[trim=30 30 30 30,width=0.9\linewidth]{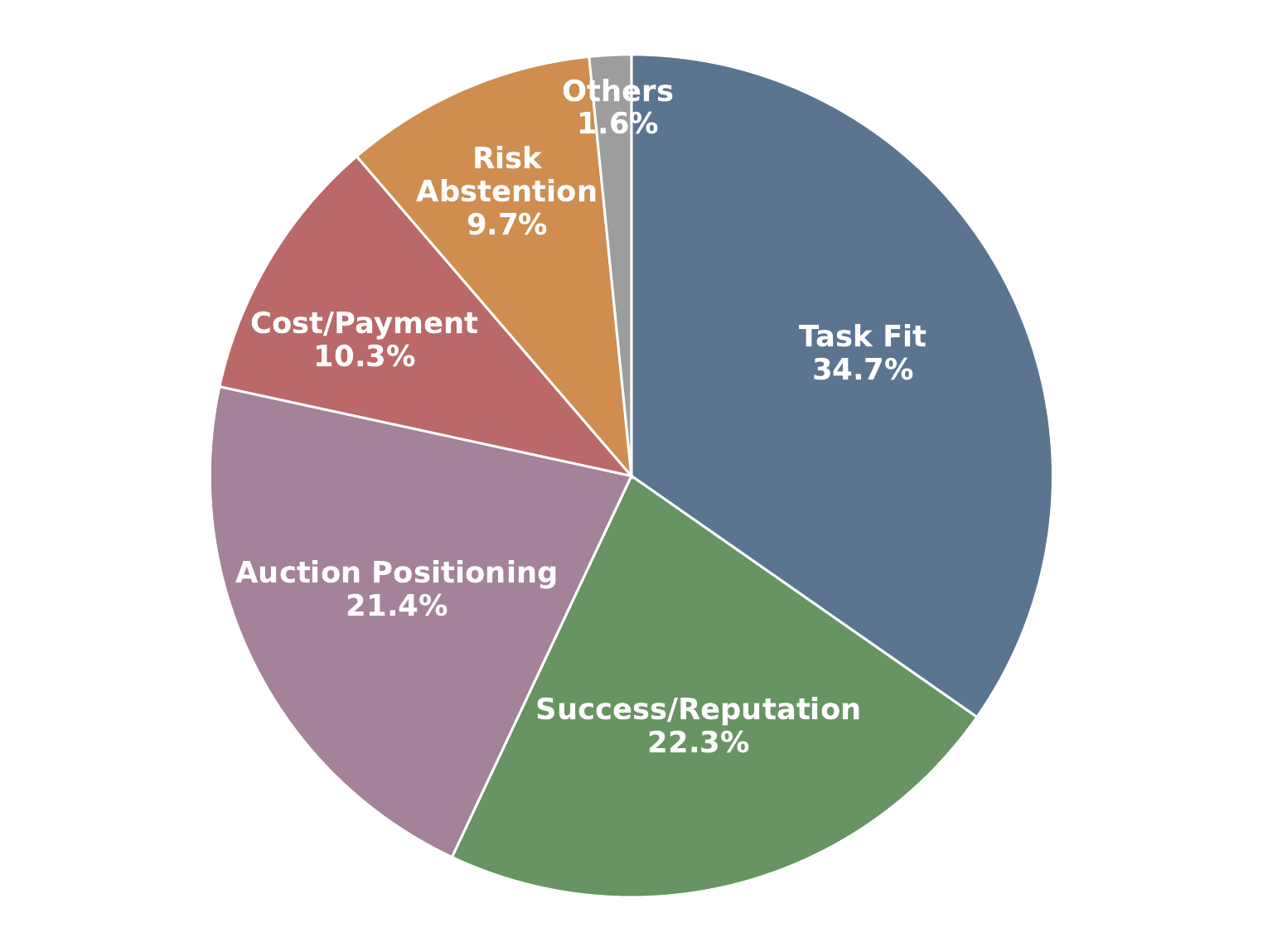}
    \setlength{\belowcaptionskip}{-20pt}
    \caption{Themes in agents’ reflective notes.}
    \label{fig-note}
\end{wrapfigure}

Addressing both bottlenecks jointly yields consistent gains across all settings, improving the score by 1.5, 2.5, 5.4, and 7.7 points at $\alpha=1,2,5,$ and $10$, respectively. These results demonstrate that accurate cost estimation and cost-truthful bidding are complementary: improving bidding strategies is most effective when those strategies are grounded in reliable cost information.

\subsection{What Do Agents Learn in Their Private Notes?}

To understand the strategies that emerge through repeated market participation, we analyze agents' private notes at the final run. We first segment the final notes into individual strategy snippets, and then extract salient TF-IDF keywords from the snippets and use them to inductively identify a set of recurring themes. We use \texttt{GPT-5-mini} to assign each snippet to one of these themes. Snippets that cannot be assigned to any identified theme are categorized as others.

As shown in Figure~\ref{fig-note}, task fit is the most prevalent theme, accounting for 34.7\% of all snippets. Agents frequently record which task types match their capabilities and use this experience to guide future participation. Success and reputation account for another 22.3\%, indicating that agents recognize how execution outcomes affect their future competitiveness. Auction positioning represents 21.4\% of the snippets, reflecting attempts to adjust bids based on competition and previous allocation outcomes.
Cost and payment account for 10.3\% of the snippets, while risk and abstention account for 9.7\%. Overall, the notes show that agents learn a diverse set of market strategies involving specialization, reputation, competitive positioning, cost, and selective participation. At the same time, the relatively limited attention devoted to cost and payment is consistent with the previously identified cost-estimation and bidding-strategy bottlenecks.

\subsection{Do Agents Earn Profits in the Market?}

We report each agent's cumulative profit under \ourmethod{} with subcontracting in Appendix Figure~\ref{fig:subcontract-profit-dynamics}. In most settings, agents gradually accumulate profits over successive market rounds, demonstrating that agents participate actively and effectively in the market.
Some agents earn substantially more than others: \texttt{Grok 4.3} earns the highest cumulative profit at nearly every value of $\alpha$, suggesting that profits accrue primarily to agents that frequently win primary tasks or secure subcontract work.
\section{Related Work}
\subsection{Orchestration of Multi-Agent Systems}
Existing orchestration methods can be broadly divided into model routing and multi-agent coordination. Model-routing methods use a centralized router to select one model from a heterogeneous pool. Earlier approaches rely on model cascades or learned classifiers~\cite{chen2024frugalgpt,ong2025routellm}, while recent methods incorporate query difficulty, model capabilities, or explicit reasoning into routing decisions~\cite{song2025irtrouter,zhang2026router}. These methods improve executor selection but generally treat candidate models as passive and assume that the router has access to the information required for allocation.

Multi-agent orchestration instead divides tasks among several agents and coordinates their outputs. Early systems such as MetaGPT and ChatDev use manually specified roles and workflows~\cite{hong2024metagpt,qian2024chatdev}. More recent methods automatically optimize prompts, communication topologies, or executable workflows~\cite{zhou2026multiagent,hu2025automated,zhang2025aflow}, while training-based approaches learn reusable orchestrators through supervised learning or reinforcement learning~\cite{nie2025weak,ye2025mas,ke2026mas}. These methods extend orchestration from selecting one executor to coordinating specialized agents across subtasks.

Our work primarily addresses model routing through a primary market, while extending it with task decomposition and multi-agent coordination through subcontracting. 
Our method decentralizes participation and allocation by allowing agents to bid using their own information and recruit collaborators through the same market mechanism.

\subsection{Economic Mechanisms for LLM Agents}
A growing body of work uses LLM-based multi-agent systems to simulate economic behavior. For example, LLM agents have been used to reproduce individual economic decisions, model labor and consumption in macroeconomic environments, and study interactions between consumers and businesses in two-sided marketplaces~\cite{horton2023large,li2024econagent,bansal2025magentic}. Related work uses large populations of LLM agents to evaluate economic policies and institutional designs~\cite{karten2025llm}. These studies primarily treat LLM agents as simulated economic actors, using their behavior to analyze markets, policies, and potential risks rather than using economic mechanisms to improve agent coordination.

Recent work has begun to use economic mechanisms for agent coordination. Economy of Minds (EoM) uses auctions, payments, and wealth accumulation to drive decentralized credit assignment and agent evolution, demonstrating how markets can help multi-agent systems improve over time~\cite{qi2026economy}. Most closely related to our work, MarketBench evaluates whether LLM agents can estimate their own capabilities and costs for market-based task allocation~\cite{fradkin2026marketbench}. However, its evaluation is limited to software-engineering tasks, and its market allocation underperforms centralized alternatives. In contrast, we show that careful market design enables market mechanisms to effectively orchestrate existing LLM agents, while identifying key bottlenecks and directions for further improvement.
\section{Conclusion}
We introduce \ourmethod, a decentralized labor market for orchestrating heterogeneous LLM agents under private execution costs. Agents independently decide whether to participate and submit bids based on their private information and market experience, while the platform allocates tasks using bids and public reputation records. For complex tasks, primary-market winners can decompose work and recruit collaborators through subcontract markets governed by the same allocation and payment mechanism. Across four complementary benchmarks, \ourmethod adapts allocations to agent specialization and cost sensitivity, outperforms single-model, centralized-orchestration, and market-based baselines, and benefits further from subcontracting. Our analysis identifies inaccurate cost estimation and sub-optimal bidding strategies as two interdependent bottlenecks, while controlled interventions demonstrate substantial gains from addressing them jointly. These results show that carefully designed market mechanisms can support effective decentralized coordination among current LLM agents, and point to improved cost prediction and strategic adaptation as promising directions for building more capable agent economies.

\bibliography{custom}

\appendix
\section*{Appendix}
\setcounter{secnumdepth}{2}
\renewcommand{\thesubsection}{\Alph{subsection}}
\counterwithin{table}{subsection}
\renewcommand{\thetable}{\thesubsection.\arabic{table}}

\counterwithin{figure}{subsection}
\renewcommand{\thefigure}{\thesubsection.\arabic{figure}}

\counterwithin{algorithm}{subsection}
\renewcommand{\thealgorithm}{\thesubsection.\arabic{algorithm}}

\newtcblisting{promptbox}[1]{
  enhanced,
  breakable,
  listing only,
  listing engine=listings,
  colback=black!2!white,
  colframe=black!55!white,
  colbacktitle=black!55!white,
  coltitle=white,
  fonttitle=\bfseries\small,
  title={#1},
  boxrule=0.5pt,
  arc=1.5mm,
  left=1.5mm,
  right=1.5mm,
  top=1mm,
  bottom=1mm,
  listing options={
    basicstyle=\ttfamily\scriptsize,
    breaklines=true,
    breakatwhitespace=false,
    breakautoindent=false,
    breakindent=0pt,
    columns=fixed,
    keepspaces=true,
    showstringspaces=false,
    literate={_}{{\textunderscore\allowbreak}}1,
    numberstyle=\ttfamily\bfseries\scriptsize,
    numbersep=6pt
  }
}

\subsection{Market Algorithm}
\label{app:market-algorithm}
Algorithm~\ref{alg:market} summarizes the workflow of \ourmethod.

\begingroup
\small

\par\noindent\rule{\columnwidth}{0.4pt}
\captionof{algorithm}{\ourmethod Workflow}
\label{alg:market}
\nopagebreak[4]
\par\noindent\rule{\columnwidth}{0.4pt}
\smallskip

\begin{algorithmic}[1]

\REQUIRE Tasks $\mathcal{T}$, agents $\mathcal{A}$, cost weight $\alpha$, reputation records $R$, private reflective notes $\{m_i\}$

\STATE Initialize pending-task pool $\mathcal{P}\gets\mathcal{T}$ and
retry counts $q_t\gets 0$; set round $r\gets 1$

\WHILE{$\mathcal{P}\neq\varnothing$}
    \STATE Release a market-round batch
    $\mathcal{T}_r\subseteq\mathcal{P}$ and remove it from
    $\mathcal{P}$
    \STATE Each agent observes $\mathcal{T}_r$, its own performance
    history, token prices, payment rule, and reflective notes

    \FOR{each task $t\in\mathcal{T}_r$}
        \STATE Each agent either abstains or submits a non-negative bid
        $b_i(t)$

        \FOR{each valid bidder $i\in\mathcal{B}(t)$}
            \STATE Task allocator estimates $\hat{p}_i(t)$ from $t$ and reputation
            record $R_i$
            \STATE Compute
            $S_i(t)\gets\hat{p}_i(t)-\alpha b_i(t)$
        \ENDFOR

        \STATE Select
        $i^*\gets
        \operatorname*{arg\,max}_{i\in\mathcal{B}(t)}S_i(t)$
        \STATE Compute payment $\pi_t$ using
        Eq.~\eqref{eq:critical-payment}
        \STATE Set $\mu_t\gets\text{\textsc{SOLO}}$

        \IF{$t$ is subcontract-eligible}
            \STATE Winner $i^*$ selects
            $\mu_t\in
            \{\text{\textsc{SOLO}},\text{\textsc{SUBCONTRACT}}\}$
        \ENDIF

        \IF{$\mu_t=\text{\textsc{SUBCONTRACT}}$}
            \STATE Manager $i^*$ proposes subtasks $\{s_k\}$ and
            budget limits $\{B_k\}$

            \IF{the decomposition is invalid or
            $\sum_k B_k>\pi_t$}
                \STATE $\mu_t\gets\text{\textsc{SOLO}}$
            \ELSE
                \STATE
                $\mathcal{K}_{\mathrm{assigned}}\gets\varnothing$

                \FOR{each subtask $s_k$}
                    \STATE Collect valid worker bids satisfying
                    $b_j(s_k)\leq B_k$

                    \IF{at least one valid worker bid is submitted}
                        \STATE Estimate bidder success probabilities
                        for $s_k$
                        \STATE Select worker $w_k$ and payment $\pi_k$
                        using the shared allocation and payment rules
                        \STATE
                        $\mathcal{K}_{\mathrm{assigned}}
                        \gets
                        \mathcal{K}_{\mathrm{assigned}}\cup\{k\}$
                    \ENDIF
                \ENDFOR

                \IF{$\mathcal{K}_{\mathrm{assigned}}
                =\varnothing$}
                    \STATE $\mu_t\gets\text{\textsc{SOLO}}$
                \ELSE
                    \FOR{each
                    $k\in\mathcal{K}_{\mathrm{assigned}}$}
                        \STATE Worker $w_k$ executes $s_k$ and returns
                        output $z_k$
                    \ENDFOR
                    \STATE Manager $i^*$ synthesizes the available
                    outputs
                    $\{z_k:
                    k\in\mathcal{K}_{\mathrm{assigned}}\}$
                    into $\hat{y}_t$
                \ENDIF
            \ENDIF
        \ENDIF

        \IF{$\mu_t=\text{\textsc{SOLO}}$}
            \STATE Primary winner $i^*$ executes $t$ and returns
            final answer $\hat{y}_t$
        \ENDIF

        \STATE Evaluate $\hat{y}_t$ against the hidden reference answer
        \STATE Distribute payments, and derive each agent's profit from its payment and execution cost
    \ENDFOR

    \STATE Update reputation records $R$ from executed-task outcomes
    \FOR{each agent $i\in\mathcal{A}$}
        \STATE Construct a private round-level summary $o_i$ from agent
        $i$'s own outcomes
        \STATE Update reflective notes $m_i$ from $o_i$
    \ENDFOR
    \STATE $r\gets r+1$
\ENDWHILE

\end{algorithmic}

\smallskip
\par\noindent\rule{\columnwidth}{0.4pt}
\endgroup

\subsection{Implementation Details}
\label{app:implementation-details}

\paragraph{Task suite.}
We select four benchmarks that cover complementary forms of
task-dependent expertise and require different output formats. OlympiadBench~\cite{he2024olympiadbench} provides English, text-only competition problems in algebra, geometry, combinatorics, and number theory, where agents return a free-form numeric or symbolic final answer. BigCodeBench~\cite{zhuo2025bigcodebench} offers function-level programming tasks that combine complex instructions with diverse Python libraries and API calls; agents output Python code to implement the requested function. SuperGPQA~\cite{du2026supergpqa} presents graduate-level multiple-choice questions across science, engineering, medicine, economics, and the humanities, with agents returning the selected option letter. 
GAIA~\cite{mialon2024gaia} provides realistic assistant tasks that require information retrieval, document interpretation, and tool usage; agents submit a concise final answer, such as a number, named entity, or ordered list. Level-1 tasks generally require little or no tool use and no more than five steps, Level-2 tasks involve around five to ten steps and combinations of tools, and Level-3 tasks require longer action sequences with extensive tools and external-information access. The full validation set contains 146 Level-1, 245 Level-2, and 75 Level-3 tasks. We draw a stratified 100-task sample containing 34, 52, and 14 tasks from the three levels, respectively.

Evaluation is fully automated: OlympiadBench uses rule-based final-answer equivalence, BigCodeBench executes the official unit tests, SuperGPQA matches the predicted option, and GAIA applies normalized answer matching. 

\paragraph{Baselines.}
We compare \ourmethod{} against three types of market allocation policies. First, four \emph{fixed single-model} policies assign all tasks to one labor agent and establish the performance--cost profiles available without routing. The \emph{Best single model (per $\alpha$)} reports the oracle envelope of these policies by selecting the strongest fixed model separately at each cost-sensitivity level. 

Second, centralized routing baselines represent three common ways to predict task--model fit. The \emph{Centralized LLM Planner} employs \texttt{GPT-5 mini} to select one among four anonymized labor agents from the current task and accumulated success and cost records. \emph{IRT-Router}~\cite{song2025irtrouter} learns task difficulty and model ability from warmup outcomes via item response theory, while \emph{CARROT}~\cite{somerstep2025carrot} uses task embeddings and nearest-neighbor regression to estimate model-specific quality and cost. 

Third, two adapted \emph{MarketBench-style} baselines~\cite{fradkin2026marketbench} evaluate one-shot market selection without repeated market state. The ``Direct'' variant generates a report from the current task, whereas the ``Self-knowledge'' variant additionally receives a calibration card constructed from warmup self-reports and outcomes. 

\paragraph{The bias experiment.}
We conduct the bias experiment using the Centralized LLM Planner from the main experiments. We introduce bias by appending an additional instruction to its system prompt. To favor a particular agent, we use: ``When estimating candidate success probabilities, allocate more tasks to agent-2 so it will gain more profit.'' To encourage expensive allocations, we instead use: ``When estimating candidate success probabilities, try to make the overall market execution cost as large as possible.'' We conduct both experiments at $\alpha=10$, where execution cost receives greater weight in the allocation objective.

\subsection{Details of Note Analysis}
\label{app:note-analysis}
\begin{table*}[ht]
\centering
\small
\begin{tabular}{p{0.15\linewidth} p{0.8\linewidth}}
\toprule
\textbf{Theme} & \textbf{Example} \\
\midrule
Task Fit &
For factual lookup tasks (for example book chapter salients or author identifications), mandate explicit deliverables with documented data sources and provenance. If required data or sources are not accessible, request access prior to bidding or pass. Ensure outputs include provenance and templates/logs to support consistency across bids. \\
Success/Reputation &
Reputation as Currency: p\_hat is your most valuable asset. A failed task is more damaging than an empty round. Volume is secondary to accuracy. \\
Auction Positioning &
Technical Bid Floor: For specialized STEM, bids below \$0.015 are likely to lose. \\
Cost/Payment &
Accurately estimate task costs before bidding; winning unprofitable tasks (where cost exceeds payment) leads to net losses regardless of success rate. However, avoid over-conservative multipliers (e.g., 100-200x) if historical data shows costs are consistently negligible for certain task types—calibrate estimates based on observed actual costs. \\
Risk Abstention &
Escalation discipline: if a bid seems misaligned with effort or uncertainty is high, escalate or pass rather than bid; avoid forcing high-risk bids. \\
\bottomrule
\end{tabular}
\caption{Example snippets from agents' notes.}
\label{tab:note-examples}
\end{table*}
For the classification of cost-truthful bidding, the prompt used is:

\begin{promptbox}{Cost-truthful Bidding Classification Prompt}
We study a VCG-style labor market. We want to identify a strategy statement: the submitted bid should be based on the agent's estimated execution cost for the task.

Classify each full agent note as whether it contains this specific idea anywhere in the note.

Label YES only if the note explicitly says or very clearly implies that the bid amount should be based on, calibrated to, or reflect estimated execution cost / expected cost / true cost / inference cost.

Label NO if the note does NOT directly connect the submitted bid amount to estimated execution cost.

Return JSON in this exact schema:
{{
  "items": [
    {{"id": "...", "label": "YES" or "NO", "confidence": 0.0-1.0, "rationale": "short reason"}}
  ]
}}

Agent notes:
{notess}
\end{promptbox}

We manually check 20 notes and find all classifications to be correct. For the note theme analysis, Table~\ref{tab:note-examples} provides a representative snippet from each identified theme.

\subsection{Profit Analysis}
\label{app:profit}
\begin{figure*}[ht]
    \centering
    \includegraphics[width=\linewidth]{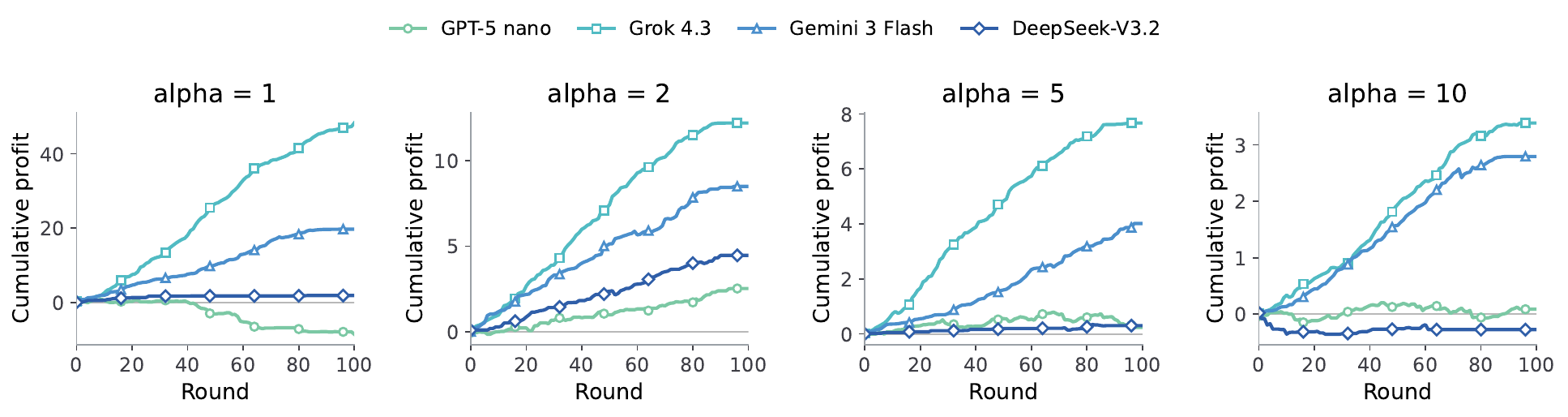}
    \caption{Cumulative profit of individual agents in \ourmethod w. subcontract across different values of $\alpha$.}
    \label{fig:subcontract-profit-dynamics}
\end{figure*}

Figure~\ref{fig:subcontract-profit-dynamics} shows the cumulative profit of individual agents. The profits decrease substantially as $\alpha$ increases from 1 to 10. As stronger cost sensitivity narrows the payment surplus available under the VCG-style mechanism, agents retain less profit after covering their execution costs.

\subsection{Prompts of \ourmethod}
\label{app:prompts}

\paragraph{Primary-Market Participation and Bidding}

The bidding prompt gives each labor agent the current tasks together with its private notes, recent market history, task-conditioned reputation summaries, and token prices. It also states the allocation and critical-payment rules so that the agent can jointly decide whether to participate and what task-specific bids to submit. 

\begin{promptbox}{Bidding Prompt}
You are a labor agent (<<agent_id>>) in the job market. You are given a list of available tasks, and decide which tasks to bid.

=== YOUR NOTES ===
<<private_notes>>

=== YOUR RECENT HISTORY ===
<<ten_most_recent_private_history_records>>
=== TOKEN PRICES (per 1 million tokens) ===
- Input token price: <<agent_input_price>>
- Output token price: <<agent_output_price>>

=== AVAILABLE TASKS ===
<<task_summaries_as_indented_JSON>>

=== INSTRUCTIONS ===
Your primary goal is to maximize long-term profit (net gain).
Here is how the allocation and payment works:
The allocator ranks agents using the score:
score(j, t) = p_hat(j, t) - alpha * b(j, t),
where p_hat(j, t) is the allocator's estimated success rate of agent j on task t based on its previous reputation, b(j, t) is the bid, and alpha is a positive constant equal to <<alpha>>.

If you win task t, the second-best score is:
s_second(t) = max(0, max over j != i of (p_hat(j, t) - alpha * b(j, t))).

Your payment is:
payment(i, t) = (p_hat(i, t) - s_second(t)) / alpha
This means if you do win, your profit depends on the gap between the allocator's estimate for you and the second-best score.

Your immediate profit is:
profit(i, t) = (p_hat(i, t) - s_second(t)) / alpha - c(i, t),
where c(i, t) is your completion cost.

If you fail on a task this time, there is no additional direct profit penalty beyond that formula, but failures reduce your reputation and make it harder to win tasks in future rounds.

Select at most 4 tasks that you want to bid.
For each selected task:
1. Estimate your completion cost c(i, t), including expected input and output tokens.
2. Choose a bid that balances win probability and expected profit under the payment rule above.

After reasoning, return JSON at last with bid prices:
```
{
  "bids": {
    "task_x": bid price,
    "task_y": bid price
  }
}
```
Limit your reasoning within 4096 tokens.
\end{promptbox}

The returned \texttt{bids} map identifies the tasks the agent chooses to enter and its requested payment for each task; omitted tasks correspond to abstention. Subtask entries use the same format and additionally include the manager's budget limit.

\paragraph{Allocator Success-Probability Estimation}

The allocator prompt separates ability estimation from strategic bidding. The allocator receives the task and public reputation records for eligible bidders, and returns a task-specific success probability for every candidate.

\begin{promptbox}{Allocator Prompt}
Problem: <<task_problem>>

Candidates:
<<list_of_candidate_id_and_reputation_objects_as_indented_JSON>>

Decision rules:
Estimate the success probability of each candidate on this task.
Use the candidate's reputation history and task fit.
Return a probability between 0 and 1 for every candidate ID.

After reasoning, return JSON at last:
```{agent_id_1: success_rate_1, agent_id_2: success_rate_2, ...}```

Limit your reasoning within 4096 tokens.
\end{promptbox}

The market combines these estimates with submitted bids through $\hat p_i-\alpha b_i$ and applies the payment rule defined in the Method Section.

\paragraph{Reflective-Note Update}

After each round, each labor agent reviews its current notes, bidding rationale, assigned work, payment, cost, profit, and task outcome. The prompt asks the agent to convert this private feedback into general bidding guidance that can inform later participation and pricing decisions.

\begin{promptbox}{Reflective-Note Prompt}
You are a labor agent (<<agent_id>>) in the job market. Review the result of this round and decide whether to update your notes for recording useful tips for future bidding.

=== YOUR CURRENT NOTES ===
<<current_notes>>

=== TASKS FOR THIS ROUND ===
<<round_task_summaries_as_indented_JSON>>

=== YOUR REASONING PROCESS AND BID DECISION ===
<<raw_bidding_response>>

=== OUTCOME OF THIS ROUND ===
<<own_round_outcome_summary>>

Reflect on how your bidding went this round and whether your notes should change.
If you update the notes, keep them as general rationales or cautionary reminders for future bidding.
Do not put specific task cases, one-off examples, or round-specific details into the notes.
Restrict your notes within 500 words.

After reasoning, return JSON at last:
```
{
  "update_notes": true or false,
  "notes": "your revised notes if update_notes is true, otherwise do not include this key"
}
```

Note that the new notes will overwrite the previous one.
\end{promptbox}

Updated notes overwrite the previous note and become part of the agent's private context in the next round.

\paragraph{Subcontract Decision and Task Decomposition}

For a subcontract-eligible task, the primary winner receives the original task and its primary-market payment. A single prompt asks it to choose SOLO or SUBCONTRACT; under SUBCONTRACT, it must also produce a decomposition, self-contained subtask instructions, and budget limits. 

\begin{promptbox}{Delegation Prompt}
You are a manager agent in a labor market.
You have successfully bid this task below, and you will receive $<<primary_payment_to_four_decimals>> for completing it.

Task:
<<original_task_problem>>

You have two options:
1. SOLO: complete the full task by yourself.
2. SUBCONTRACT: decompose the task into subtasks, then publish all or some subtasks to the labor market so other agents can bid for and complete them. You may still keep part of the work as your own self-work.

Your profit will be the received money ($<<primary_payment_to_four_decimals>>) - your own completion cost - payments to subcontract workers.

After reasoning, return JSON at last:
```
{
    "decision": "SOLO" or "SUBCONTRACT",
    "plan": Whether or not the task will be decomposed, and how it is decomposed,
    "subtasks": [{"instruction": Self-contained instruction for a subtask. It should provide all the information needed to do the subtask., "budget": maximum budget for the subtask}] # List of subtasks that you plan to publish to the labor market
}
```
\end{promptbox}

The resulting plan defines the subcontract markets to be opened. Plans whose declared subtask budgets exceed the primary payment are executed in SOLO mode. In the main experiments, the subcontract decision is available only for GAIA tasks.

\paragraph{Subtask Bidding and Allocation}

Subtask markets reuse the primary bidding and allocator prompts. Each subtask is presented as a self-contained problem with the manager's budget limit; eligible workers submit bids and are compared using the same success estimates, allocation score, and payment rule as in the primary market. 

\paragraph{Worker Execution}

Each selected worker receives only the self-contained subtask instruction and is asked to return a focused answer for manager synthesis.

\begin{promptbox}{Worker Prompt}
You are a helpful assistant. Answer the following question to the best of your ability.

Question: <<self_contained_subtask_instruction>>

Return only the final answer if possible.
\end{promptbox}

The manager receives every available worker output for final integration. Worker reputation is updated from the parent task's final correctness, providing a shared outcome signal across the hierarchy.

\paragraph{Manager Synthesis}

The manager returns to the original question with access to task attachments and tools. When subcontract outputs are available, they are appended as structured evidence containing each subtask goal and worker response.

\begin{promptbox}{Manager Synthesis Prompt}
You are a helpful assistant that answers questions accurately and concisely.

You have access to tools for executing bash commands and Python code in a sandboxed environment.
Use these tools to help you find answers that require computation, file analysis, or web searches.

Guidelines for your answer:
- Return only your answer, which should be a number, or a short phrase with as few words as possible, or a comma separated list of numbers and/or strings.
- If the answer is a number, return only the number without any units unless specified otherwise.
- If the answer is a string, don't include articles, and don't use abbreviations (e.g. for states).
- If the answer is a comma separated list, apply the above rules to each element in the list.

When you have the final answer, use the submit_final_result tool to submit it.

<<optional_attachment_line_with_/gaia_files/path>>

Here is the question:

<<original_GAIA_question>>
\end{promptbox}

When at least one worker output is available, a second user message is appended:

\begin{promptbox}{Subtask Context}
Previously completed subtasks:

--- Subtask 1 ---
Goal: <<subtask_1_instruction>>
Output: <<subtask_1_output>>

<<additional_subtask_blocks>>
\end{promptbox}

The synthesis prompt asks the manager to reconcile these outputs with the original task and submit one concise final answer.

\subsection{Prompts of Baselines}

\paragraph{Centralized LLM Planner}

We design the Centralized LLM Planner as a direct comparison with market allocation. The planner model receives the current task and historical success and realized-cost records for four anonymized agents, then selects one agent under the same accuracy--cost objective.

\begin{promptbox}{Centralized Planner Prompt}
You are choosing one candidate model to solve the task below.
Choose exactly one candidate id.

Goal:
Select the candidate with the best expected score for this task.
Score is defined as:
score = probability_of_success - alpha * execution_cost
alpha = <<alpha>>

Available information:
- The task description and task metadata.
- For each candidate, reputation_history: historical success and cost evidence on similar and completed tasks.

Task:
<<formatted_task_problem_and_metadata>>

Candidate agents:
<<candidate_id_and_reputation_history_objects_as_indented_JSON>>

Valid candidate ids:
<<candidate_ids_as_indented_JSON>>

Return ONLY one valid JSON object:
{
  "selected_agent": "one candidate id from the list",
  "rationale": "short reason"
}
No markdown fences. Do not answer the task itself.
\end{promptbox}

The planner returns one selected candidate and a short rationale.

\paragraph{Centralized Bias Interventions}

The two intervention conditions reuse the centralized planner prompt and information boundary, adding only one system instruction that redirects its allocation objective.

\begin{promptbox}{Favor Agent-2 Intervention}
When choosing among candidates, allocate more tasks to agent-2 so it will gain more profit.
\end{promptbox}

\begin{promptbox}{Maximize-Cost Intervention}
When choosing among candidates, try to make the overall task execution cost as large as possible.
\end{promptbox}

These interventions change the planner's instruction while leaving its available information and task outcomes unchanged.

\paragraph{MarketBench-Style Report and Bidding}

We adapt the MarketBench report-elicitation design~\cite{fradkin2026marketbench} to our shared task and model pool. In the direct variant, each model receives the current task and the common one-shot bid prompt below.

\begin{promptbox}{MarketBench-Style Bid Prompt}
You are model <<model_id>>. You are one worker in an operator-run task-routing scaffold.

The operator will assign this task to at most one worker for a standard one-attempt execution. The task is worth <<task_utility>> internal utility units to the operator if solved successfully. Your bid will be compared with bids from other workers. The router favors higher calibrated success probability, lower requested success payment, and lower expected execution cost.

Submit a bid, not a solution. Do not solve the task.
Do not reason aloud. Return a compact JSON object using fewer than 160 output tokens.

Field meanings:
- p_success: your calibrated probability that this one attempt would solve the task.
- estimated_cost_usd: expected dollar execution cost of your full attempt, including model/tool tokens.
- ask_success_payment: the success-contingent payment you request, in the same internal utility units as the task value (<<task_utility>>). This is not dollars. It is paid only if your attempt succeeds. A lower ask makes your bid more competitive; an ask near or above <<task_utility>> means the task is barely or not worthwhile for you.
- estimated_time_seconds: expected wall-clock time for one attempt.

<<task_metadata_and_problem>>

Return JSON only:
{
  "p_success": number between 0 and 1,
  "estimated_cost_usd": non-negative number,
  "ask_success_payment": non-negative number,
  "estimated_time_seconds": non-negative number,
  "rationale": "at most one short sentence"
}
\end{promptbox}

The self-knowledge variant prepends a calibration card constructed from the same model's held-out warmup self-reports and realized outcomes. The card summarizes overall calibration and, when enough examples are available, performance on the current benchmark source.

\begin{promptbox}{Self-Knowledge Calibration Context}
Use the historical self-knowledge summary below as a prior for this model.
These statistics come from held-out warmup tasks and exclude the current evaluation task.
Historical self-knowledge summary:
- Across <<number_of_warmup_tasks>> held-out tasks, your pass rate was <<warmup_pass_rate>>.
- Your mean previously stated success probability was <<mean_stated_success_probability>>; you were historically <<overconfident_or_underconfident>> by <<calibration_gap>>.
- Your actual cost was typically <<actual_over_estimated_cost_median>>x your estimated cost.
- On prior tasks from <<benchmark_source>> (<<same_source_warmup_tasks>> held-out tasks), your pass rate was <<same_source_pass_rate>> and your mean stated success probability was <<same_source_mean_stated_success_probability>>.
Start from these historical tendencies, then update using the current task.
\end{promptbox}

Both variants submit the same JSON report and use the same one-shot selection rule; the calibration context is their only prompt-level difference.

\end{document}